\documentclass[11pt,twoside]{article}  
\usepackage{asp2006}
\usepackage{adassconf}

\begin{document}   

%
%

\paperID{P3.2}

%

\title{A User Interface for Semantically Oriented Data Mining of Astronomy Repositories}

%
%
%
%
%

\markboth{User Interface for Semantic Data Mining}{Thomas \it{et al.}}

%
%
%
%

\author{Brian Thomas and Edward Shaya}
\affil{University of Maryland, Department of Astronomy, College Park, MD 20742.}

%

\contact{Brian Thomas}
\email{thomas@astro.umd.edu}

%
%
%

\paindex{Thomas, B.}
\aindex{Shaya, E.}     

%



\begin{abstract}          
We present a user-friendly, but powerful interface for the data mining of 
scientific repositories. We present the tool in use with actual astronomy
data and show how it may be used to achieve many different types of powerful
semantic queries. The tool itself hides the gory details of query formulation, 
and data retrieval from the user, and allows the user to create workflows
which may be used to transform the data into a convenient form.
\end{abstract}

%
%

\section{Motivation and Introduction}
Our goal is to create a tool which a scientist may use to author sophisticated queries 
for data from astronomy repositories and then apply a variety of operations to these 
data so that the data of interest are acquired. The concept is straightforward. The user 
indicates the "source" data they want by creating a class with restrictions on its 
properties. Operations are then applied to convert data into the "goal" form. A directional 
graph (or workflow) indicates the order and precedence of operations. 

For this work, we have focused on application of some semantic technologies. In particular,
our tool makes use of an underlying OWL [1] ontology to understand the relationships between
astronomical objects and their properties. Using this knowledge our tool may aid the user to
prevent creating illegal queries and to aid in the selection of appropriate properties and 
applicable operations when designing their workflow. 
The tool also handles the conversion of the source class definition into a SPARQL query [2] 
and manages the interaction with the astronomical repository (which is a SPARQL endpoint).

\section{Viper Graphical User Interface}

\begin{figure}[t]
\epsscale{0.70}
\plotfiddle{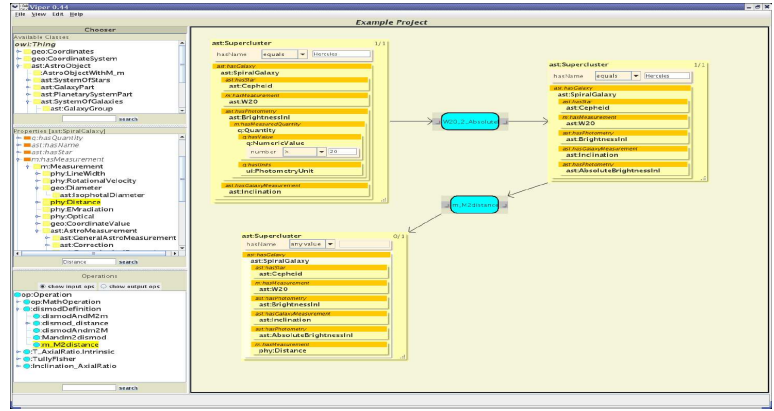}{200pt}{270.0}{40.0}{30.0}{-150pt}{180pt}
\caption{A screen shot of an example project using Viper.} \label{P3.2-fig-1}
\end{figure}

The Viper GUI (figure~\ref{P3.2-fig-1}) is partitioned into 2 main areas, a “Chooser”, 
on the left side which is comprised of a number of "palettes" which control how the
user may design their workflow and a workspace area, on the right side, where items 
in the chooser may be dropped to create the science workflow. The workflow shown in
figure~\ref{P3.2-fig-1} describes a search for spiral galaxy data from the Hercules 
supercluster. The spiral galaxy data should have at least one Cepheid star, as well as 
the properties of W20, Inclination and I-band brightness (apparent magnitude). The
last property is also constrained to be brighter than 20th magnitude. 

\begin{figure}[t]
\plotfiddle{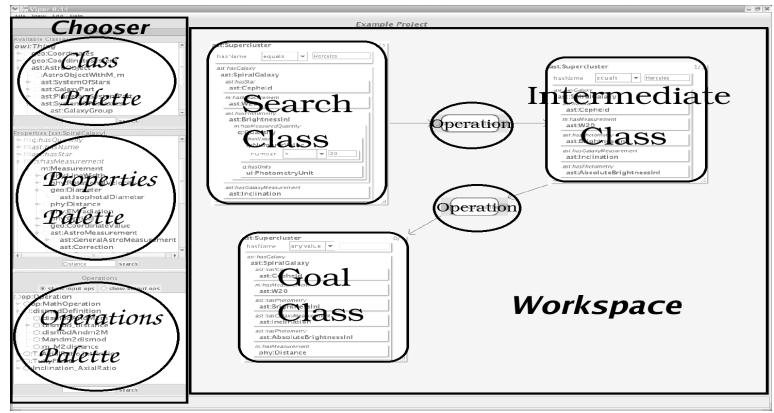}{200pt}{270.0}{40.0}{30.0}{-150pt}{180pt}
\caption{Diagram of significant areas in tool.} \label{P3.2-fig-2}
\end{figure}

Details of the Chooser and Workspace areas are discussed below. 
Figure~\ref{P3.2-fig-2} is a guide to the more prominent features of the tool.

\subsection{Chooser}
The chooser is divided vertically into 3 sections, the "Class Palette", the "Properties Palette" 
and the "Operations Palette". The Class Palette selection may be dragged and dropped into the
workspace to create classes in the workflow. The available properties in the Properties Palette 
are restricted to whatever class is selected in the Class Palette, hiding illegal properties.
Properties from the tree may be dragged over to the selected class to add them 
\footnote{In the search target class only.} and may be further restricted according to their
type.
The Operations Palette shows legal operations for the selected Class, depending on whether the
user wishes to see operations which have the selected class as input or output. 

\subsection{Workspace}

The workflow is a linear chain of alternating classes and operations and starts with the definition 
of the type of science data to retrieve, the “search target” and the last class is the "goal". 
Classes (boxes) from the Chooser are dropped on the workspace. 
For the selected class in the workspace, properties may then be added.
Some properties are object properties, and in this case further recursion is possible where the 
object of a property may itself have properties (looks like nested boxes in the workspace). 
Datatype properties may constrain simple data such as strings or numerical values.
Operations appear in the workspace as linking circles. The characteristics of the operations may
be toggled/changed as appropriate (not shown).

\begin{figure}[t]
\plotfiddle{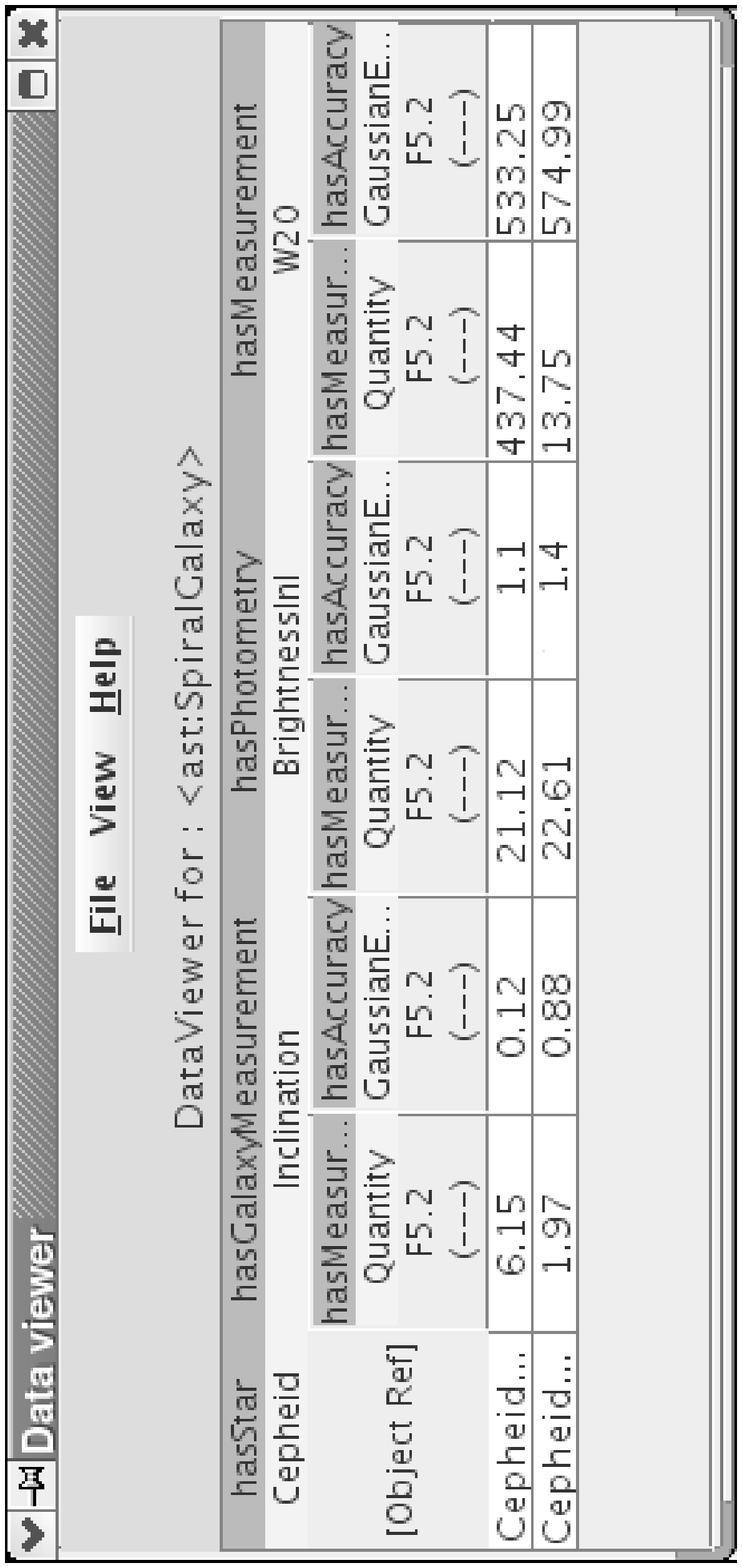}{200pt}{270.0}{40.0}{40.0}{-150pt}{200pt}
\caption{Screen shot of table view of instance data.} \label{P3.2-fig-3}
\end{figure}

The numbers in the north east corner of the classes represent the number of instances 
which are available at repositories (right hand number), and the number which are 
"available" (left hand side\footnote{in the search target this means the number downloaded,
in other classes it means how many instances are present at that step}).
Available instances are viewable, in flattened form, as tables (figure~\ref{P3.2-fig-3}) for any
selected class in the workflow.

Operations are the only way to change the composition of properties in a class which is not 
the search target. In figure~\ref{P3.2-fig-1} the middle class in the workflow 
shows a new property, the I-band absolute magnitude, has been added 
in a child property class (SpiralGalaxy) after applying a Tully-Fisher transformation.

\section{Summary}

We have been successful at creating an initial easy to use GUI to manage access to 
semantic repositories. In order to be applied, the GUI requires that the repository
utilize a framework ontology to define measurements and operations [3]. Otherwise, the
archivist is free to design any hierarchy of classes they may choose, and software
such as D2R [4] makes the mapping of existing tabular databases into a SPARQL endpoint
fairly easy. 

Nevertheless, there remains significant work to be done on Viper. Presently the tool
only manages an interaction with a single repository. We hope that this may be expanded,
so as to include multiple repositories, and perhaps interaction to discover repositories
via a registry. In terms of improved usability, we hope
to investigate how inference might allow a user to specify the goal class first,
and work backwards to various source targets.
Furthermore, we expect that by adding special operations to allow the union and intersection
of classes will allow for useful non-linear workflows.

\end{document}